\documentclass[12pt,english]{revtex4-1}
\usepackage[T1]{fontenc}
\usepackage[latin9]{inputenc}
\usepackage{verbatim}
\usepackage{amsmath}
\usepackage{amssymb}

\makeatletter
\allowdisplaybreaks

\makeatother

\usepackage{babel}
\begin{document}
\title{Transient Dynamics from Quantum to Classical\\
- From the Developed Coherent State via Extreme Squeezing -}
\author{Masahiro Morikawa}
\email{hiro@phys.ocha.ac.jp}

\affiliation{Department of Physics, Ochanomizu University, 2-1-1 Otsuka, Bunkyo,
Tokyo 112-8610, Japan}
\begin{abstract}
We explore the transient dynamics associated with the emergence of
the classical signal in the full quantum system. We start our study
from the instability which promotes the squeezing of the quantum system.
This is often interpreted as the particle production though being
reversible in time. We associate this state a non-dissipative classical
fluctuations and study their trigger to develop the coherent state
which can be classical if sufficiently developed. The Schwinger-Keldysh
in-in formalism yields the classical Langevin equation including the
fluctuation force which faithfully reflects the particle production
property of the original quantum system. This formalism is applied
to some transient process; the initiation of the spontaneous symmetry
breaking, appearance of the off-diagonal long-range order in Bose-Einstein
condensation, a transient process of the classicalization of the quantum
fluctuations in the inflationary cosmology,... and gives some implications
on the origin of the irreversibility associated with the transition
from quantum to classical. 
\end{abstract}
\maketitle

\section{introduction}

Quantum mechanics is no doubt a complete theory and correctly describes
nature in the most fields in laboratory science. On the other hand,
in some situations when the operational control doesn't work, quantum
mechanics seems to be lacking any autonomous description. This applies
to the problem of the origin of inhomogeneity in the early Universe,
the black hole information loss problem, and so on. All of them seem
to be related to the emergence, in quantum mechanics, of any classical
signal which spontaneously violates the original symmetry of the system. 

We pickup three typical examples. 

i) In the early Universe, the standard cosmology indicates that the
ultimate origin of the spatially non-uniform density fluctuations
$\delta\rho(t,\boldsymbol{x})$ is the quantum fluctuations of the
inflaton field $\delta\hat{\phi}(t,\boldsymbol{x})$ in the inflationary
stage when the Universe transforms from microscopic to macroscopic
scales through the exponential expansion. There is no explicit observer
nor detector in the early Universe and therefore quantum mechanics
cannot yield any definite outcome of the quantum fluctuations. In
order to solve this problem, some authors introduce detectors \citep{Nambu2009}\citep{Matsumura2018}or
modify the basic laws of quantum mechanics so that the wave function
spontaneously collapses \citep{Simonov2016}. This situation is similar
to the quantum-information-theory aspect of the black hole \citep{Hawking1976}.
If one tries to describe the black hole and the Hawking radiation
as a unitary evolution, one encounters the violation of the equivalence
principle\citep{Almheiri2013}. 

ii) A presice description of the spontaneous symmetry breaking (SSB)
introduces the two limiting operations, large volume and small external
field limits, which are carefully ordered. However, this is an explicit
vilation of the symmetry that the violation direction is artificially
determined from the begining. On the other hand, in the Bose-Einstein
condensation(BEC), the classical order parameter $\varphi_{0}$ describes
the amount of condensation of boson gas which accumulates into the
lowest energy state. This is also the indicator how extent the phase
symmetry is broken spontaneously. In the ordinary argument, this classical
value is simply assumed from the beginning as simply dividing the
total quantum field $\hat{\Phi}(t,\boldsymbol{x})=\hat{\phi}(t,\boldsymbol{x})+\varphi_{0}$
\citep{Lieb2007}. However, the trasient time development from 0 to
$\varphi_{0}$ is actually needed to describe the initiation of BEC
process. 

iii) Furthermore, we still do not have a full successful model of
the quantum measurement apparatus, in which a definite classical signal
$m(t)$ emerges that has a firm correlation with the further measurement
of the same quantum system\citep{Leggett2005}. If we adopt the standard
operational prescription of quantum mechanics, the measurement can
be fully described by POVM. However, from the fundamental level, the
measurement apparatus should be described by quantum mechanics as
well as the system. In this case, the classical signal can have many
values before the completion of the measurement. However, after the
measurement it must choose one value among such possible values. This
process is not simply the SSB above but must further include the back
reaction process to the system\citep{Morikawa2006}. 

All the above examples have three main characteristics, beside the
appearance of the macroscopic classical signal. 

1) \textit{state selection}: The system chooses one state among many
other possible states. If this all states can be transformed with
each other by a local symmetry operation, then this is exactly the
process of the spontaneous symmetry breaking. 

2) \textit{classical statistical probability}: At this stage, deterministic
evolution is terminated but an intrinsic probability appears which
governs the fate of the system. 

3) \textit{arrow of time}: This process is intrinsically irreversible.
The system can evolve into any one state among many leaving a classical
signal. However, this signal cannot be canceled and the system cannot
return to the original state which allows another possibility to develop. 

In order to describe the transient process of the development from
quantum to classical yielding the classical signal, we follow the
following two steps in this paper. 

a) \textit{squeezed state}: If the quantum state is unstable, say
by the negative mass squared, the system tends to evolve into the
squeezed state. This state allows classical statistical description,
in the effective action method, which is well separated from the unitary
dynamical evolution. However, at the free level without any interaction,
this statistical fluctuation, being non-dissipative, never appears
outside. At this stage, everything is still reversible and unitary
whatever the squeezing is strong. Furthermore the original symmetry,
if any, is not yet broken. The statistical fluctuations, all superposed,
are always symmetric. 

b) \textit{coherent state}: If the above free system couples to the
other state or the non linear interaction appears, then the above
statistical fluctuations do affect the full system. If this coherent
state develops either by the strong statistical fluctuations or by
strong instability of the potential, the classical portion of the
coherent state dominates the quantum portion. Thus the classical signal
appears possibly with dissipation. 

In this way, the classicality thus finally obtained in b) is a quantitative
notion; there is a continuous transition from quantum to classical.
On the other hand, the classicality of the noise in a) is a qualitative
notion; all the fluctuations are classical. 

We follow these two steps in the subsequent sections. In section 2,
we study the appearance of the squeezed state in the unstable system
and show that this squeezing process can be identified as the particle
production which can be interpreted as the classical statistical fluctuations
withoiut dissipation. In section 3, we consider the non-linearity
of the system in which this fluctuation triggers the development of
the coherent state. In section 4, we motivate the introduction of
the Schwinger-Keldysh in-in formalism starting from the popular Langevin
equation. In section 5, we introduce the in-in effective action method
to describe the above two processes, a) and b), at once. In section
6, we briefly describe some relevant examples which are directly related
with our formalism. In the last section 7, we summarize our study. 

\section{Squeezing - non dissipative dry noise}

An instability or the time-dependent classical source yield the squeezed
state in general. We first consider a simple model which yields the
squeezed state, the inverted harmonic oscillator\citep{Albrecht1994}.
The Hamiltonian is given by 

\begin{equation}
\hat{H}=\frac{1}{2m}\hat{p}^{2}-\frac{m\omega^{2}}{2}\hat{q}^{2}=i\frac{\mathrm{\omega\hslash}}{2}\left(\hat{a}^{2}e^{-2i\phi}-h.c.\right),\label{eq:HinvertedHO}
\end{equation}
where 
\begin{equation}
\begin{array}{cc}
\hat{a}^{\dagger}= & (\frac{m\omega}{2\hbar})^{1/2}(q-i\frac{p}{m\omega})\\
\hat{a}= & (\frac{m\omega}{2\hbar})^{1/2}(q+i\frac{p}{m\omega})
\end{array},
\end{equation}
 and $\phi=-\pi/4.$ The ordinary cross term $\hat{a}^{\dagger}\hat{a}$
disappears because the second term in the middle of Eq.(\ref{eq:HinvertedHO})
is negative. Then the wave function at time $t$ becomes 

\begin{align}
\left.|\Psi\left(t\right)\right\rangle  & =\exp\left[\frac{\omega t}{2}\left(\hat{a}^{2}e^{-2i\phi}-h.c.\right)\right]\left.|0\right\rangle \equiv S(t)\left.|0\right\rangle \\
 & =\exp\left[-\frac{\omega t}{2}\left(\hat{a}^{\dagger}\right)^{2}e^{2i\phi}-\left(\frac{\omega t}{2}\right)^{2}\right]\left.|0\right\rangle .
\end{align}
Therefore, in the state $\left.|\Psi\left(t\right)\right\rangle $,
the particle pair is 'condensed'. The operator $S(t)$ defines the
Bogolubov transformation from the canonical pair ${a,a^{\dagger}}$
to the new pair :

\begin{equation}
\left\{ \begin{array}{l}
b=S^{\dagger}aS=\hat{a}\cosh\omega t-\hat{a}^{\dagger}e^{2i\phi}\sinh\omega t,\\
b^{\dagger}=S^{\dagger}a^{\dagger}S=\hat{a}^{\dagger}\cosh\omega t-\hat{a}e^{-2i\phi}\sinh\omega t,
\end{array}\right.\label{eq:Bogolubov1}
\end{equation}
 and $SS^{\dagger}=S^{\dagger}S=1$. This state $\left.|\Psi\left(t\right)\right\rangle $
is unlimitedly squeezed in time toward the direction $\phi=-\pi/4$
in phase space as,

\begin{equation}
\left\langle \Psi\left(t\right)|\right.\left(\hat{p}\cos\phi\pm\hat{q}\sin\phi\right)\left.|\Psi\left(t\right)\right\rangle =\left\{ \begin{array}{l}
4e^{-2t}\\
4e^{2t}
\end{array}\right..
\end{equation}

This state is often regarded as the particle-creating state as 
\begin{equation}
N\equiv\left\langle \Psi\left(t\right)|\right.a^{\dagger}a\left.|\Psi\left(t\right)\right\rangle =\left\langle 0|\right.b^{\dagger}b\left.|0\right\rangle =(\sinh\omega t)^{2}.
\end{equation}
However, $\left.|\Psi\left(t\right)\right\rangle $ is simply a quantum
mechanical state and the particles claimed to be created are clearly
not the classical object before any observation. Actually, the state
can be reversible to the original state $\left.|0\right\rangle $
by the operation $S^{\dagger}$. This means that this inverted harmonic
oscillator, if prepared in the symmetric state initially, never fall
to some particular direction. The state is always in the symmetric
neutral position $\left\langle \Psi\left(t\right)|\right.\hat{x}\left.|\Psi\left(t\right)\right\rangle =0$
even if the quantum fluctuations develop infinitely. On the other
hand, since the kinetic energy $K$ and the negative of the potential
energy $-V$ both ever increases, the action $A=K-V$ of the system
would diverge. This may indicate some classical property of this system
$A\gg\hbar$ \citep{Albrecht1994}. Since all the one point function
vanishes, unique property of the state appears in the two point functions:
\begin{equation}
\begin{array}{cc}
\left\langle \left[x\left(t\right),x\left(t'\right)\right]\right\rangle  & =\frac{i}{2}\left(\frac{m\omega}{2\hbar}\right)^{-1}\sinh\left(\omega\left(t-t'\right)\right),\\
\left\langle \left\{ x\left(t\right),x\left(t'\right)\right\} \right\rangle  & =\frac{1}{2}\left(\frac{m\omega}{2\hbar}\right)^{-1}\cosh\left(\omega\left(t+t'\right)\right),
\end{array}\label{eq:2ptfunctions}
\end{equation}
where the former is locally ($t\approx t'$) normal but the latter
abnormally diverge in time. 

More generally in quantum field theory, the Hamiltonian Eq.(\ref{eq:HinvertedHO})
is the infinite collection of the harmonic oscillator labeled by the
three momentum $\mathbf{k}.$ Since the momentum is conserved, the
pair in the squeezed state must have exactly the opposite momentum.
Thus the Bogolubov transformation Eq.(\ref{eq:Bogolubov1}) should
now be the form,
\begin{equation}
\left\{ \begin{array}{l}
b_{\mathbf{k}}=\alpha_{\mathbf{k}}^{*}\hat{a}_{\mathbf{k}}-\beta\hat{a}_{\mathbf{-k}}^{\dagger},\\
b_{\mathbf{-k}}^{\dagger}=\alpha_{\mathbf{k}}\hat{a}_{-\mathbf{k}}^{\dagger}-\beta_{\mathbf{k}}^{*}\hat{a}_{\mathbf{k}}
\end{array}\right..\label{eq:Bogolubov2}
\end{equation}
Therefore the particle pair of momentum $\mathbf{k},\mathbf{-k}$
is entangled with each other. Many cases of the particle production,
Unruh effect, accelerated mirror, Hawking radiation from the black
hole,...\citep{Birrell1982} correspond to this type of Bogolubov
transformation and the states are slightly generalized squeezed states.
These are simply the pure quantum states represented by the wave functions.
These states are also similar to the entangled spin pair of Silver
atoms in the Stern\textendash Gerlach experiments \citep{Gerlach1922}.
All the above quantum system, even under the external force, is free
and therefore keeps quantum coherence, and is even reversible, until
observed or disturbed by the quantum interaction. 

\section{Coherent state -development of classicality}

We have introduced the squeezed state in the previous section in relation
with quantum-classical transition. We have an another popular state,
the coherent state, which is worth consideration in our context. The
coherent state is define to be the eigenstate of the annihilation
operator, 
\begin{equation}
\hat{a}\left|\alpha\right\rangle =\alpha\left|\alpha\right\rangle 
\end{equation}
which has an explicit form,
\begin{align}
\left|\alpha\right\rangle  & =e^{\alpha\hat{a}^{\dagger}-\alpha^{*}\hat{a}}\left|0\right\rangle \\
 & \equiv C\left(\alpha\right)\left|0\right\rangle \\
 & =e^{-\frac{\left|\alpha\right|^{2}}{2}}e^{\alpha\hat{a}^{\dagger}}\left|0\right\rangle .
\end{align}
The operator $C\left(\alpha\right)$ shifts the creation and annihilation
operators by C-number
\begin{equation}
\left\{ \begin{array}{l}
\hat{b}=C^{\dagger}\hat{a}C=\hat{a}+\alpha,\\
\hat{b}^{\dagger}=C^{\dagger}\hat{a}^{\dagger}C=\hat{a}^{\dagger}+\alpha^{*}.
\end{array}\right.\label{eq:Ceffect}
\end{equation}

The particle number expectation is 
\begin{equation}
N=\left\langle \alpha\right|\hat{a}^{\dagger}\hat{a}\left|\alpha\right\rangle =\left|\alpha\right|^{2}
\end{equation}
and the two developed coherent states, \emph{i.e.} individually large
$\alpha,\beta$, have almost no superposition 
\begin{equation}
\left|\left\langle \alpha\right.|\left.\beta\right\rangle \right|^{2}=e^{-\left|\alpha-\beta\right|^{2}},
\end{equation}
in particular, for the infinite system, the volume factor $V$ enhances
this property. 
\begin{equation}
\propto e^{-V\left|\alpha-\beta\right|^{2}}.
\end{equation}

The coherent state is not special but is generally produced by the
time dependent single variable interaction $\xi\left(t\right)\widehat{x}$,
where $\xi\left(t\right)$ is a classical external variable. This
is contrasted with the squeezed state which is generally produced
by the time dependent square variable interaction $m\left(t\right)\widehat{x}{}^{2}$.
For example, let us consider the quantum system is exerted by the
classical field $\xi(t)$, 
\begin{equation}
\ddot{\widehat{x}}(t)=-\gamma\dot{\widehat{x}}(t)-V'(\widehat{x}(t))+\xi(t).\label{eq:Langevinquantum}
\end{equation}
Then the general solution would be 
\begin{equation}
\widehat{x}(t)=\widehat{x}(0)+\int^{t}dt'\triangle\left(t-t'\right)\xi\left(t'\right)\label{eq:qu+cl}
\end{equation}
where the second term on the RHS, including any appropriate green
function $\triangle\left(t-t'\right)$, represents the accumulation
of the classical field. Thus, in this case, 
\begin{equation}
\widehat{x}(t)=C\left(t\right)^{\dagger}\widehat{x}\left(0\right)C\left(t\right)
\end{equation}
where
\begin{equation}
C\left(t\right)=\exp\left[\int^{t}dt'\triangle\left(t-t'\right)\xi\left(t'\right)\widehat{x}\left(0\right)\right].
\end{equation}
Thus the time evolution induced by the single variable interaction
$\xi\left(t\right)\widehat{x}$ yields a coherent state. If the classical
part gradually dominates in Eq.(\ref{eq:qu+cl}) due to the accumulation
of the external force $\xi\left(t\right)$, then the state gradually
becomes more classical. 

It would be useful to comment that the external classical field $\xi\left(t\right)$
needs not to be a systematic force, but can be random with zero-mean.
Even in this case, the classical component increases in time and develop
classicality. 

\section{Langevin to in-in formalism - a motivation}

In order to consider the transient process from quantum to classical,
let us consider a simple example of the Langevin equation first. This
gives us a natural motivation to the Schwinger-Keldysh in-in closed
path time quantum theory (CTP)\citep{Kita2010}. Starting from the
classical Langevin equation, we derive the effective partition function.
The quantization of this partition function almost deduces the CTP
formalism in quantum mechanics. 

The Langevin equation is a dynamical equation of a particle in the
environment with the potential force $-V'$ and the random force $\xi$
with friction $\gamma$ : 
\begin{equation}
\ddot{x}(t)=-\gamma\dot{x}(t)-V'(x(t))+\xi(t).\label{eq:Langevinoriginal-1}
\end{equation}
The statistical average 
\begin{equation}
\left\langle ...\right\rangle _{\xi}=\int D[\xi]...P[\xi]
\end{equation}
is determined by the Gaussian weight functional $P[\xi]$,
\begin{equation}
P[\xi]=e^{-\int\xi(t)^{2}/(2\sigma^{2})}.\label{eq:Pgiven}
\end{equation}
We would like to know the action which drives this Langevin equation.
We first try to construct the partition function of the system. The
partition function is derived by summing all the possible motions
in the whole phase space.

\begin{eqnarray}
Z[J] & \equiv & \left\langle \delta[\ddot{x}(t)+\gamma\dot{x}(t)+V'(x(t))-\xi(t)]\right\rangle _{\xi}\\
 & = & \int D[\xi]P[\xi]\delta[\ddot{x}(t)+\gamma\dot{x}(t)+V'(x(t))-\xi(t)]\nonumber \\
 & = & \int D[\xi]D[x']P[\xi]e^{i\int dtx'(t)\{\ddot{x}(t)+\gamma\dot{x}(t)+V'(x(t))-\xi(t)\}}\nonumber \\
 & = & \int D[x']e^{i\tilde{S}[x,x']}\nonumber 
\end{eqnarray}
where the integral form of the delta functional is utilized introducing
a fictitious variable $x'\left(t\right)$, and the action becomes
\begin{equation}
\tilde{S}[x,x']\equiv\int dt\{-\dot{x}'(t)\dot{x}(t)+\gamma x'(t)\dot{x}(t)+x'(t)V'(x(t))+\frac{i}{2}\sigma^{2}x'(t)^{2}\}\label{eq:Langevin action}
\end{equation}
where the boundary term is dropped. Note that the last term, which
represents classical statistical fluctuations, is pure imaginary in
the action. The rest of the terms describe the deterministic dynamics
though including frictional term. 

We now quantize this system because the most basic theory would be
the quantum mechanics, from which classical dynamics eventually appear.
The partition function for the quantized system simply becomes

\begin{equation}
Z[J]=\int D[x]D[x']e^{i\tilde{S}[x,x']}\label{eq:PI Langevin}
\end{equation}
in the path integral form. The mixed expression of the two variables
$x\left(t\right)$and $x'\left(t\right)$ in Eq.(\ref{eq:Langevin action})
is dowdy. It is possible to rewrite the action more resemble to the
ordinary dynamics. In order to do so, we rewrite the variables as
\begin{equation}
x_{\pm}=x\pm\frac{1}{2}x'.
\end{equation}
Then, Eq. (\ref{eq:PI Langevin}) becomes
\begin{equation}
Z[J]=\int D[x_{+}]D[x_{-}]e^{i\tilde{S}[x_{+},x_{-}]}
\end{equation}
where
\begin{equation}
\tilde{S}[x_{+},x_{-}]=\int dt\left\{ \begin{array}{cc}
\left(\left(\ddot{x}_{+}\left(t\right)\right)^{2}-V(x_{+}(t))\right)-\left(\left(\ddot{x}_{-}\left(t\right)\right)^{2}-V(x_{-}(t))\right)\\
+\frac{\gamma}{2}\left(x_{+}(t)\dot{x}_{-}(t)-\dot{x}_{+}(t)x_{-}(t)\right)+\frac{i}{2}\sigma^{2}\left(x_{+}(t)-x_{-}(t)\right)^{2}
\end{array}\right\} .\label{eq:Sderived}
\end{equation}
The first line of the above represents the deterministic dynamics
for the variables $x_{\pm}\left(t\right)$ separately, and the second
line dissipation and fluctuation terms where the variables $x_{\pm}\left(t\right)$
mix up. It is a natural extension of this expression to introduce
a closed time-path $C$: which starts from $-\infty$ to $\infty$
($+$ branch) and then comes back from $\infty$ to $-\infty$ ($-$
branch). We suppose the supports of the variables $x_{\pm}\left(t\right)$
are, respectively, the $+$ and $-$ branches. We denote the variable
$\tilde{x}\left(t\right)$ on the countour $C$ unifying the variables
$x_{\pm}\left(t\right)$ :
\begin{equation}
\tilde{x}\left(t\right)=\begin{cases}
x_{+}(t) & t\in(+\mathrm{branch})\\
x_{-}(t) & t\in(-\mathrm{branch}).
\end{cases}
\end{equation}
We use this unification for all the variables and denote them by tilde. 

It is possible to reverse the above logic to get to the action $\tilde{S}[x_{+},x_{-}]$;
starting from the action $\tilde{S}[x_{+},x_{-}]$ to get to the Langevin
equation. If we find a complex action including an extra degrees of
freedom like $x'$ in the above, classical random field appears and
the evolution equation becomes the Langevin equation. We study a typical
case in the next section. 

\section{CTP to Langevin}

The best way to describe the transient process from quantum to classical
would be the Schwinger-Keldysh in-in formalism\citep{Kita2010}. In
this theory, the partition function for the system with the free action 

\[
S[x]=\int dt(\dot{x}^{2}+\omega^{2}x^{2}),
\]
is given by 

\[
\begin{array}{cc}
\tilde{Z}[\tilde{J}] & =\int_{C}D\tilde{x}\exp[iS[\tilde{x}]+i\int dt\tilde{J}(t)\tilde{x}(t)]\equiv\exp i\tilde{W}\\
= & \int D\tilde{x}\exp[iS[x_{+}]-iS[x_{-}]+i\int dt\tilde{J}(t)\tilde{x}(t)],
\end{array}
\]
where the tilde denotes the variables on the closed-path $C$ as before.
This reduces to 
\begin{equation}
\tilde{Z}[\tilde{J}]=\exp[-\frac{1}{2}\int dt\tilde{J}(t)\tilde{G}_{0}(t,t')\tilde{J}(t')].\label{eq:Zfree}
\end{equation}
where
\begin{equation}
\tilde{G}_{0}(t,t')=\left(\begin{array}{ll}
G_{F}(t,t') & G_{+}(t,t')\\
G_{-}(t,t') & G_{\overline{F}}(t,t')
\end{array}\right)\equiv\left(\begin{array}{ll}
\mathrm{Tr}\left[Tx(t)x(t')\rho\right] & \mathrm{Tr}\left[x(t')x(t)\rho\right]\\
\mathrm{Tr}\left[x(t)x(t')\rho\right] & \mathrm{Tr}\left[\overline{T}x(t)x(t')\rho\right]
\end{array}\right)
\end{equation}
where $T$ denotes the ordinary time-ordering operator and $\overline{T}$
the anti time-ordering operator. If we change the representation of
the matrix by
\begin{equation}
J_{\pm}(t)=J_{c}\pm\frac{1}{2}J_{\Delta}
\end{equation}
or
\begin{equation}
\tilde{J}=\left(\begin{array}{cc}
J_{\Delta}\\
J_{C}
\end{array}\right)=\left(\begin{array}{cc}
1 & -1\\
\frac{1}{2} & \frac{1}{2}
\end{array}\right)\left(\begin{array}{cc}
J_{+}\\
J_{-}
\end{array}\right),
\end{equation}
then, we have 
\begin{equation}
\tilde{G}_{0}(t,t')=\left(\begin{array}{ll}
0 & G_{R}(t,t')\\
G_{A}(t,t') & G_{C}(t,t')
\end{array}\right)\equiv\left(\begin{array}{ll}
0 & \theta(t-t')\mathrm{Tr}[x(t'),x(t)]\rho\\
\theta(t'-t)\mathrm{Tr}[x(t'),x(t)]\rho & \mathrm{Tr}\{x(t),x(t')\}\rho
\end{array}\right).
\end{equation}

In our case, 
\begin{equation}
\begin{array}{cc}
G_{R}(t,t') & =\theta(t-t')\left\langle \left[x\left(t\right),x\left(t'\right)\right]\right\rangle \\
 & =\frac{i}{2}\left(\frac{m\omega}{2\hbar}\right)^{-1}\sin\left(2\phi\right)\theta(t-t')\sinh\left(\omega\left(t-t'\right)\right)\\
 & =\frac{i}{2}\left(\frac{m\omega}{2\hbar}\right)^{-1}\theta(t-t')\sinh\left(\omega\left(t-t'\right)\right),
\end{array}\label{eq:Gr}
\end{equation}
\begin{equation}
\begin{array}{cc}
G_{C}(t,t') & =\left\langle \left\{ x\left(t\right),x\left(t'\right)\right\} \right\rangle \\
 & =\frac{1}{2}\left(\frac{m\omega}{2\hbar}\right)^{-1}\left(\cosh\left(\omega\left(t+t'\right)\right)-\cos\left(2\phi\right)\sinh\left(\omega\left(t+t'\right)\right)\right)\\
 & =\frac{1}{2}\left(\frac{m\omega}{2\hbar}\right)^{-1}\cosh\left(\omega\left(t+t'\right)\right),
\end{array}\label{eq:Gc}
\end{equation}
where $\phi=-\pi/4$. 

These two types of green functions sometimes show different infrared
behavior. Actually, for low frequency $\omega\left(t-t'\right)\ll1$
and $\omega\left(t+t'\right)\ll1$, 
\begin{equation}
\begin{array}{cc}
G_{R}(t,t')\propto\sinh\left(\omega\left(t-t'\right)\right)\rightarrow\omega\left(t-t'\right),\\
G_{C}(t,t')\propto\cosh\left(\omega\left(t+t'\right)\right)\rightarrow1.
\end{array}
\end{equation}
$G_{R}(t,t')$ has milder IR behavior than $G_{C}(t,t')$ by the factor
$\omega$. However, as we will see shortly, $G_{C}(t,t')$ can be
separated from the action which describes the deterministic dynamics.
In particular, this situation becomes prominent at the inflationary
stage in the early Universe\citep{Morikawa2016}. 

Comparing Eqs.(\ref{eq:Gr},\ref{eq:Gc}), the symmetric term in Eq.(\ref{eq:Zfree})
\begin{equation}
-\frac{1}{2}\int dtJ_{\Delta}(t)G_{C}(t,t')J_{\Delta}(t')
\end{equation}
is real and positive. Therefore we can factor out this part as classical
statistical fluctuations introducing an auxiliary field $\xi\left(t\right),$
\begin{equation}
\tilde{Z}[\tilde{J}]=\int D\xi P\left(\xi\right)\exp[-\frac{1}{2}\int dt\tilde{J}(t)\tilde{G}'_{0}(t,t')\tilde{J}(t')+i\int dtJ_{\Delta}(t)\xi(t)],\label{eq:Zderived}
\end{equation}
where 
\begin{equation}
P\left(\xi\right)=\exp[-\frac{1}{2}\int dt\xi(t)G_{C}(t,t')\xi(t')],\label{eq:Pderived}
\end{equation}
and $\tilde{G}'_{0}(t,t')$ is thus separated green function. This
separation procedure is just a reverse of the previous section where
Eq.(\ref{eq:Sderived}) yields Eq.(\ref{eq:Pgiven}). 

As in the above, we have identified the classical statistical fluctuation,
\textit{i.e.} noise, in Eqs.(\ref{eq:Zderived}, \ref{eq:Pderived}).
However, this noise only couples to the field $J_{\Delta}(t)$, in
Eq.(\ref{eq:Zderived}), which is an external source term but not
any dynamical variable. Therefore this noise never comes out as is.
Nothing happens in free state before any measurement process or interaction
according to the quantum mechanics theory. Furthermore, the noise
here is dry and is not accompanied by dissipation. 

This situation drastically changes if we introduce the interactions.
Let us introduce the effective action, in the in-in formalism, which
is the Legendre transformation of the partition function $\tilde{Z}[\tilde{J}]=\exp i[\tilde{W}[\tilde{J}]]$,
\begin{equation}
\exp[i\tilde{\Gamma}[\tilde{X}]]=\exp i[\tilde{W}[\tilde{J}]-\int dt\tilde{J}(t)\tilde{X}(t)].
\end{equation}
This becomes 
\begin{eqnarray}
\exp[i\tilde{\Gamma}[\tilde{X}]] & = & \exp i[\tilde{W}[\tilde{J}]-\int dt\tilde{J}(t)\tilde{X}(t)]\nonumber \\
 & = & \int_{C}D\tilde{x}\exp i[\tilde{S}[x]+\int d^{4}x\tilde{J}(x)(\tilde{x}(t)-\tilde{X}(t))]\nonumber \\
 & = & \int_{C}D\tilde{x}\exp i[\tilde{S}[\tilde{X}+\tilde{x}]+\int dt\tilde{J}(t)\tilde{X}(t)],
\end{eqnarray}
where we shifted the path-integration variable. Expanding $\tilde{S}[\tilde{X}+\tilde{x}]$
in the series of $\tilde{x}$, we have

\begin{equation}
\tilde{S}[\tilde{X}+\tilde{x}]=\tilde{S}\left[\tilde{X}\right]+\tilde{S}^{\prime}\left[\tilde{X}\right]\tilde{x}+\frac{1}{2}\tilde{S}''\left[\tilde{X}\right]\tilde{x}^{2}+\frac{1}{3!}\tilde{S}'''\left[\tilde{X}\right]\tilde{x}^{3}+\ldots.\label{eq:Sseries}
\end{equation}
The first term represents the action for the classical field $\tilde{X}\left(t\right),$the
second term and the third term yield interaction for the quantum variable,
and the second term a free action for $\tilde{x}$ in the background
of $\tilde{X}$. 

If the interaction is quartic, $\lambda x\left(t\right)^{4},$ and
the background $\tilde{X}=0$ initially, then the term $\frac{1}{3!}\tilde{S}'''\left[\tilde{X}\right]\tilde{x}^{3}$
gives the dominant interaction term $\lambda\tilde{X}\left(t\right)\tilde{x}\left(t\right)^{3}$.
The lowest contribution of this term yields the two-loop quantum effect
for $\tilde{X}\left(t\right)$
\begin{align}
 & \int dtdt'\lambda^{2}\tilde{X}\left(t\right)\mathrm{Tr}\left[T_{C}\rho\tilde{x}\left(t\right)^{3}\tilde{x}\left(t'\right)^{3}\right]\tilde{X}\left(t'\right)\\
= & \int dtdt'\lambda^{2}\tilde{X}\left(t\right)\mathrm{Tr}\left[T_{C}\rho\tilde{x}\left(t\right)\tilde{x}\left(t'\right)\right]^{3}\tilde{X}\left(t'\right)\\
= & \int dtdt'\lambda^{2}\tilde{X}\left(t\right)G_{C}\left(t,t'\right)^{3}\tilde{X}\left(t'\right)\\
= & \int dtdt'\lambda^{2}\left(X_{C},X_{\Delta}\right)_{t}\left(\begin{array}{cc}
0 & G_{C}^{2}\text{\ensuremath{G_{A}}}\\
\text{\ensuremath{G_{R}}}\text{\ensuremath{G_{C}}}^{2} & \text{\ensuremath{G_{C}}}^{3}
\end{array}\right)_{t,t'}\left(\begin{array}{cc}
 & X_{C}\\
 & X_{\Delta}
\end{array}\right)_{t'}
\end{align}
Therefore, the imaginary term $X_{\Delta}\left(t\right)G_{C}\left(t,t'\right)^{3}X_{\Delta}\left(t\right)$
contributes to the fluctuation as before and can be separated from
the real part of the action by the introduction of the auxiliary field
$\xi\left(t\right)$, 

We have, in the lowest order, 
\begin{equation}
\exp[i\Gamma[X]]=\int D\xi P\left(\xi\right)\exp[iS_{\mathrm{eff}}[X]],
\end{equation}
where the real action is
\begin{equation}
\begin{array}{cc}
S_{\mathrm{eff}}[X]= & S[X]+\int\int dtdt'X_{C}(t)\left(1+\lambda^{2}G_{C}\left(t,t'\right)^{2}\right)G_{A}\left(t,t'\right)X_{\Delta}(t')\\
 & +\int\int dtdt'X_{\Delta}(t)G_{R}\left(t,t'\right)\left(1+\lambda^{2}G_{C}\left(t,t'\right)^{2}\right)X_{C}(t')+\int dt\xi(t)X_{\Delta}(t)],
\end{array}
\end{equation}
and the fluctuation weight is given by
\begin{equation}
P\left(\xi\right)=\exp[-\frac{\lambda^{2}}{2}\int dt\xi(t)G_{C}\left(t,t'\right)^{3}\xi(t')].\label{eq:P3}
\end{equation}
Note that the advanced term $\int\int dtdt'X_{C}(t)\left(1+\lambda^{2}G_{C}\left(t,t'\right)^{2}\right)G_{A}\left(t,t'\right)X_{\Delta}(t')$
yields the same retarded term if we exchange the variables $t\leftrightarrow t'$. 

Now the application of the variational principle for $S_{\mathrm{eff}}[X]$
\begin{equation}
\frac{\delta S_{\mathrm{eff}}[X]}{X_{\Delta}(t)}|_{X_{\Delta}=0}=0,
\end{equation}
yields the classical Langevin equation as 
\begin{equation}
\ddot{X}_{C}(t)-\omega^{2}X_{C}(t)+2\int dt'G_{R}\left(t,t'\right)\left(1+\lambda^{2}G_{C}\left(t,t'\right)^{2}\right)X_{C}(t')+\xi(t)=0.\label{eq:generalLangevin}
\end{equation}

This equation describes the rapid evolution of the classical variable
$X_{C}(t)$ under a) the strong fluctuations $\xi(t)$ with Eqs.(\ref{eq:P3},\ref{eq:Gc})
violently disturb the system, and b) the original classical instability
$-\omega^{2}X_{C}(t)$ which promotes the exponential development
of the system. The retarded term $2\int dt'G_{R}\left(t,t'\right)\left(1+\lambda^{2}G_{C}\left(t,t'\right)^{2}\right)X_{C}(t')$
sometimes shows the dissipative effects. 

The above equation (\ref{eq:generalLangevin}) describes the evolution
of the classical variable $X_{C}(t)$ from 0, the symmetric state,
to a finite value, the asymmetric state. Eventually in this evolution,
other interaction terms in Eq.(\ref{eq:Sseries}) gradually contribute.
Some of them yield the new type of noise. Individual imaginary term
seems to yield individual random noise. 
\begin{equation}
\exp[-X_{\Delta}G_{C}X_{\Delta}]=\int D\xi_{1}\exp[-\xi_{1}G_{C}^{-1}\xi_{1}+i\xi_{1}X_{\Delta}],
\end{equation}
and
\begin{equation}
\exp[-X_{\Delta}G_{C}^{2}X_{\Delta}]=\int D\xi_{2}\exp[-\xi_{2}G_{C}^{-2}\xi_{2}+i\xi_{2}X_{\Delta}].
\end{equation}
However, all the perturbation terms have interference with each other
and therefore should be treated at once: 
\begin{equation}
\exp[-\lambda^{2}X_{\Delta}\left(G_{C}+G_{C}^{2}+G_{C}^{3}\right)X_{\Delta}]=\int D\xi\exp[-\xi\left[\lambda^{2}\left(G_{C}+G_{C}^{2}+G_{C}^{3}\right)\right]^{-1}\xi+i\xi X_{\Delta}].
\end{equation}
Since the fluctuation kernel $\lambda^{2}X_{\Delta}\left(G_{C}+G_{C}^{2}+G_{C}^{3}\right)X_{\Delta}$
is unique, the random field $\xi(t)$ is also unique; random fields
do not appear separately. 

\section{Some Applications and Implications}

Our argument is general and will have many applications and implications.
Some of them are very briefly described below. Individual argument
will be reported separately. 

\subsection{Transient dynamics of the spontaneous symmetry breaking and the Bose-Einstein
condensation }

A standard method to describe the Spontaneous symmetry breaking (SSB)
needs an infinitesimal explicit violation of the symmetry with the
delicate order of the two limiting operations. For example in the
case of ferromagnetic materials, the full order parameter is given
by 
\begin{equation}
M_{\pm}\equiv\lim_{B\rightarrow0\pm}\lim_{V\rightarrow\infty}m_{V}\left(B\right),\label{eq:doublelimit}
\end{equation}
where $m_{V}\left(B\right)$ is the local average of the magnetization\citep{Lieb2007}.
In the case of Bose-Einstein condensation (BEC), the argument starts
from the assumption that the boson field can be separated as \citep{Pitaevskii2003}
\begin{equation}
\hat{\phi}=\left\langle \hat{\phi}\right\rangle +\hat{\delta\phi},\label{eq:assumedseparation}
\end{equation}
where the classical parameter $\left\langle \hat{\phi}\right\rangle $
represents the 0-momentum condensation. 

On the other hand, based on our approach, general SSB can be described
as follows. We suppose the unstable potential for the complex scalar
field $\phi\left(x\right)$ such as 
\begin{equation}
V\left(\phi\right)=\frac{1}{2}m^{2}\phi^{2}+\frac{1}{4!}\lambda\phi^{4}
\end{equation}
with $m^{2}<0$ and $\lambda>0$. The initial tachyonic instability
around $\phi\approx0$ induces the squeezed state and thus dry noise.
Through the interaction, this noise autonomously violates the $U\left(1\right)$
symmetry and allow the development of the classical degrees of freedom
as a coherent state. Therefore, we do not need Eq(\ref{eq:doublelimit})
and SSB is genuinely spontaneous. Furthermore, the artificial separation
is not needed as in Eq.(\ref{eq:assumedseparation}) and the field
$\left\langle \hat{\phi}\right\rangle $ does emerge as a developed
coherent state. 

In the case of BEC, the condensation is not only characterized by
\begin{equation}
\frac{\left\langle \hat{a}_{0}^{*}\hat{a}_{0}\right\rangle }{V}>0,
\end{equation}
but also needs the condition\citep{Lieb2007} 
\begin{equation}
\frac{\left|\left\langle \hat{a}_{0}\right\rangle \right|^{2}}{V}>0.
\end{equation}
which guarantees the off-diagonal long-range order or the fact that
the BEC as a phase transition accompanying the spontaneous symmetry
breaking. In BEC, the Gross-Pitaevskii equation is generally given
as Eq.(\ref{eq:generalLangevin}). The fluctuation term disappears
when the system moves out from the unstable transient region ($\left|\left\langle \hat{\phi}\right\rangle \right|^{2}>-2m^{2}/\lambda$). 

\subsection{Transient dynamics of the quantum measurement}

The appearance of the classicality is deeply related with the quantum
measurement process in which a particular state is probabilistically
selected among multiple possibilities. The situation is very similar
to SSB above, however, there must be a back-reaction to the quantum
system from the emerged classical degrees of freedom. This back-reaction
guarantees the firm correlation between the quantum system and the
measurement apparatus. 

The prototype of the quantum measurement model has been analyzed in
this line of thought introducing the external thermal bath in\citep{Morikawa2006}.
This model describes the transient dynamics of the detector field
$\hat{\phi}$ measures the spin $\hat{\mathrm{S}}.$ The Lagrangian
is given by 
\begin{equation}
L=\frac{1}{2}\left(\nabla\hat{\phi}\right)^{2}-\frac{1}{2}m^{2}\hat{\phi}^{2}-\frac{1}{4!}\lambda\hat{\phi}^{4}+\mu\hat{\phi}\boldsymbol{\mathrm{\hat{S}B}}+(bath).
\end{equation}
From the present approach, the detector should be described by $X_{C}(t)$
in Eq.(\ref{eq:generalLangevin}). On top of this dynamics, the back
reaction of it to the spin $\hat{\mathrm{S}}$ is needed. The thermal
bath degrees of freedom will not be needed. The fluctuation would
be provided by the dry noise associated to the initial squeezed state
triggered by the unstable potential ($m^{2}<0$ and $\mu>0$). 

\subsection{Transient dynamics which shows macroscopic irreversibility}

A classical degrees of freedom as a developed coherent state has appeared
after the time evolution by the Langevin equation. This process cannot
be canceled and is irreversible. This is true even if the energy dissipation
does not exist. Actually, the Brownian motion of the classical degrees
of freedom is described by the Langevin equation with random fluctuations.
The recursion probability, the system comes back to the original position,
would be vanishingly small after the Brownian motion described by
the Langevin equation. 

In the careful argument on the appearance of the arrow of time in
quantum mechanics \citep{Ordonez2017}, the essence of the irreversibility
is the appearance of the decaying and growing pair, as well as the
natural boundary condition which picks up one from the pair. The first
essence is similar to our tachyonic modes and the second part is automatically
selected by the Schwinger-Keldysh in-in formalism. Actually we use
it to put the retarded and advanced contribution together in the evolution
equation(\ref{eq:generalLangevin}). We have further discussed the
appearance of the classical degrees of freedom. 

\subsection{Transient dynamics from quantum to classical fluctuations in the
inflationary cosmology}

In the cosmology, the hypothetical scalar field inflaton is supposed
to cause the inflation in the early Universe and also to yield the
ultimate seeds of density fluctuations \citep{Weinberg2008}. This
inflaton field and the space-time metric mix together to yield the
gauge invariant variables. We consider such scalar variable $v_{\mathbf{k}}\left(\eta\right)$
where $\mathbf{k}$is the three momentum and $\eta$$\left(\in\left(-\infty,0\right)\right)$
is the conformal time variable. The Hamiltonian for $v_{\mathbf{k}}\left(\eta\right)$
in the inflation (de Sitter space) becomes \citep{Albrecht1994},
\begin{align}
H & =\frac{1}{2}\int\frac{d^{3}k}{(2\pi)^{3/2}}\left[\hat{p}_{-\mathbf{k}}\hat{p}_{\mathbf{k}}+k^{2}\hat{v}_{-\mathbf{k}}\hat{v}_{\mathbf{k}}+\frac{1}{\eta}\left(\hat{p}_{-\mathbf{k}}\hat{v}_{\mathbf{k}}+\hat{v}_{-\mathbf{k}}\hat{p}_{\mathbf{k}}\right)\right]\\
 & =\frac{1}{2}\int\frac{d^{3}k}{(2\pi)^{3/2}}\left[k\left(\hat{a}_{\mathbf{k}}^{\dagger}\hat{a}_{\mathbf{k}}+\hat{a}_{-\mathbf{k}}^{\dagger}\hat{a}_{-\mathbf{k}}+1\right)+\frac{i}{\eta}\left(e^{-2i\vartheta}\hat{a}_{-\mathbf{k}}\hat{a}_{\mathbf{k}}-h.c.\right)\right]\label{eq:AlbrechtH}\\
\nonumber 
\end{align}
where $\vartheta=-\pi/2$ and the factor $1/\left|\eta\right|$ in
this Eq.(\ref{eq:AlbrechtH}) infinitely increases and yields strong
squeezed state. However this is a special quantum state and does not
directly mean the appearance of the classical fluctuations as we have
already studied. Introducing the non-linear interactions of $v$ and
constructing the effective action, we can decompose the action into
the deterministic part and the stochastic part\citep{Morikawa2016},
\begin{align}
\exp[i\widetilde{\Gamma}[\tilde{\varphi}]] & =\int{\cal D\xi}P(\xi)\exp[iS_{0}[\tilde{\varphi;\xi}]],\\
\exp[iS_{0}[\tilde{\varphi;\xi}]] & =\exp[iS_{0}[\overset{\text{\~{ }}}{\varphi}]]\int D\phi\exp i[(\lambda\varphi(x)^{3})_{\triangle}G_{R}(x-y)(\lambda\varphi(y)^{3})_{C}+\\
 & (\lambda\varphi(x)^{3})_{C}G_{A}(x-y)(\lambda\varphi(y)^{3})_{\triangle}+i(\lambda\varphi(x)^{3})_{\triangle}G_{C}(x-y)(\lambda\varphi(y)^{3})_{\triangle}]\\
P(\xi) & =\exp[-\frac{1}{4}\int d^{3}k\xi(\overrightarrow{k})G_{C}(\overrightarrow{k})^{-1}\xi(\overrightarrow{k})].
\end{align}
reflecting the dry noise generated by the squeezed state from Eq.(\ref{eq:AlbrechtH}).
The fluctuation kernel is given by

\begin{equation}
G_{C}(\overset{\rightarrow}{k})=\frac{H^{2}}{k^{3}}\left((1+k^{2}\eta\eta^{\prime})\cos(k\eta)+k\eta\sin(k\eta)\right)
\end{equation}
where $H$ is the Hubble constant for the inflationary de Sitter space. 

The Langevin equation for the field 
\begin{equation}
3H\dot{\varphi}{}_{k}+(\lambda/2)\varphi_{0}^{2}\varphi_{k}=(\lambda/2)\varphi_{0}^{2}\xi_{k}
\end{equation}
yields the classical statistical power spectrum.
\begin{equation}
\left\langle \varphi_{k}\varphi_{k}\right\rangle _{\xi}\thickapprox\lambda^{2}\varphi_{0}^{4}\frac{H^{2}}{k^{3}}.
\end{equation}
This transient process from quantum to classical has been made possible
both by the sqeezed state and the interaction of the inflaton field. 

\section{Conclusions}

We have explored the transient emergent process of the classical degrees
of freedom in the full quantum system. In this process, the existence
of the squeezing state and the development of the coherent state are
both essential. The former squeezed state may be triggered by an instability
and yields tachyonic mode. The latter coherent state develops in the
Langevin dynamics triggered by the interactions. Thus both processes
are indispensable. 

We first considered the generation of the squeezed state in the inverted
harmonic oscillator model. The degree of squeezing increases unboundedly
in time. Though the action and the expectation value of the particle
number increases and therefore one may tend to think that many particles
are actually produced. However, this is simply a squeezed state fully
described by the quantum theory and can never be interpreted as classical
nor random statistical. We have shown that some non-dissipative noise,
\textit{i.e.} dry noise, is associated to this state using the Schwinger-Keldysh
in-in formalism. At this stage, the noise never destroys the original
symmetry if any. Moreover, the generation process of this state is
reversible and the squeezing can be canceled. This is always true
for free quantum system.

Then, introducing the interaction, we considered the development of
the coherent state triggered by the above dry noise. This process
is described by the Langevin equation. The degree of the coherent
state increases, firstly by the ever accumulating dry noise and secondly
by the unstable potential. This means that the state is dominated
by the classical part which determines the finite vacuum expectation
values. 

Thus we have considered two types of general classicality in our argument.
The first one is associated to the squeezed state and is represented
by the real exponent in the in-in path-integral formalism. The second
one is associated to the development of the coherent state and is
represented by the development of the vacuum expectation values. This
latter one is the objective classicality we wanted to obtain. 

The appearance of the classicality is the key feature in many important
transient processes in Physics. They are the dynamics of the general
spontaneous symmetry breaking, the initiation process of the Bose-Einstein
condensation, the transient process of the quantum measurement and
the appearance of the classical signal in the apparatus, the transient
process that the arrow of time appears in the macroscopic system,
the transient process of the inflation in the early Universe where
the quantum fluctuations form the classical seeds of galaxies, and
so on. 
\begin{acknowledgments}
The author would like to thank all the members of the Astrophysics
group in Ochanomizu University for useful and critical comments as
well as for inspiring thorough discussions. 
\end{acknowledgments}


\begin{thebibliography}{10}
\bibitem{Nambu2009}Y. Nambu and Y. Ohsumi, Phys.Rev.D\textbf{80},124031
(2009); Phys.Rev.D\textbf{84}, 044028 (2011).

\bibitem{Matsumura2018}Akira Matsumura and Yasusada Nambu, Phys.Rev.D\textbf{98},
025004 (2018).

\bibitem{Simonov2016}K. Simonov, B.C. Hiesmayr, Phys.Rev.A\textbf{94},
052128 (2016) .

\bibitem{Hawking1976}S. W. Hawking, Phys. Rev. D\textbf{14}, 2460
(1976).

\bibitem{Almheiri2013}A. Almheiri, D. Marolf, J. Polchinski and J.
Sully, JHEP\textbf{1302}, 062 (2013). 

\bibitem{Lieb2007}Elliott H. Lieb, Robert Seiringer, Jakob Yngvason,
Rep. Math. Phys.\textbf{59}, 389-399 (2007).

\bibitem{Leggett2005}A. J. Leggett, Science\textbf{ 307}-5711, 871-872
(2005).

\bibitem{Morikawa2006}Masahiro Morikawa and Akika Nakamichi, Prog.
Theor. Phys.\textbf{116}, 2006, 679\textendash 698. (2006).

\bibitem{Albrecht1994}Andreas Albrecht, Pedro Ferreira, Michael Joyce,
Tomislav Prokopec, Phys.Rev.D\textbf{50}, 4807-4820 (1994).

\bibitem{Birrell1982}N.D.Birrell and P.C.W.Davies, \emph{Quantum
Fields in Curved Space}, Cambridge Monographs on Mathematical Physics,
Cambridge University Press (1982).

\bibitem{Gerlach1922}Gerlach, W.and Stern, O., Zeitschrift für Physik.
\textbf{9}, 349\textendash 352 (1922).

\bibitem{Kita2010}T. Kita, Prog. Theor. Phys. \textbf{123}, 581 (2010).

\bibitem{Morikawa2016}Masahiro Morikawa, arXiv:1604.01015 {[}hep-th{]}
(2016). 

\bibitem{Pitaevskii2003}L. P. Pitaevskii and S. Stringari, \textit{Bose-Einstein
Condensation} International Series of Monographs on Physics, Clarendon
Pr (2003).

\bibitem{Ordonez2017}Gonzalo Ordonez, Naomichi Hatano, J. Phys. A:
Math. Theor.\textbf{ 50,} 405304 (2017).

\bibitem{Weinberg2008}Steven Weinberg, \emph{Cosmology}, Oxford Univ
Pr (2008). 
\end{thebibliography}
\end{document}